\documentclass[aps,prb,twocolumn,superscriptaddress,showpacs]{revtex4}
\usepackage{graphicx}

\begin{document}


\title{Interplay between mesoscopic phase separation and bulk magnetism in the layered Na$_{x}$CoO$_{2}$}


\author{A. Zorkovsk\'{a}}
\affiliation{Centre of Low Temperature Physics, P. J. \v{S}af\'{a}rik University \& Slovak Academy of Science\\Park Angelinum 9, 04154 Ko\v{s}ice, Slovakia}
\author{M. Orend\'{a}\v{c}}
\affiliation{Centre of Low Temperature Physics, P. J. \v{S}af\'{a}rik University \& Slovak Academy of Science\\Park Angelinum 9, 04154 Ko\v{s}ice, Slovakia}
\author{J. \v{S}ebek}
\affiliation{Institute of Physics AS CR, Na Slovance 2, 18221 Prague, The Czech Republic}
\author{E. \v{S}antav\'{a}}
\affiliation{Institute of Physics AS CR, Na Slovance 2, 18221 Prague, The Czech Republic}
\author{P. Svoboda}
\affiliation{Department of Electronic Structures, Faculty of Mathematics and Physics, ChU, Ke Karlovu 5, CZ 12116 Prague 2, The Czech Republic}
\author{I. Bradari\'{c}}
\affiliation{The "Vin\v{c}a" Institute of Nuclear Sciences, P.O. Box 522, 11001 Belgrade, Serbia and Montenegro}
\author{I. Savi\'{c}}
\affiliation{Faculty of Physics, University of Belgrade, Studentski trg 12-14, 11000 Belgrade, Serbia and Montenegro}
\author{A. Feher}
\affiliation{Centre of Low Temperature Physics, P. J. \v{S}af\'{a}rik University \& Slovak Academy of Science\\Park Angelinum 9, 04154 Ko\v{s}ice, Slovakia}


\date{\today}

\begin{abstract}
Specific heat of the layered Na$_{x}$CoO$_{2}$ (x=0.65, 0.70 and 0.75) oxides has been measured in the temperature range of 3-360 K and magnetic field of 0 and 9 T. The analysis of data, assuming the combined effect of inter-layer superexchange and the phase separation into mesoscopic magnetic domains with localized spins embedded in a matrix with itinerant electronic character, suggests that the dominant contribution to the specific heat in the region of short-range ordering is mediated by quasi-2D antiferromagnetic clusters, perpendicular to the CoO$_{2}$ layers. 
\end{abstract}



\maketitle

The recently discovered superconductivity \cite{Takada}, and the unique combination of high thermopower with low electrical resistivity \cite{Wang} brought the sodium cobalt oxide Na$_{x}$CoO$_{2}$ and its hydrated counterpart into the focus of interest of the last years. Theoretical considerations \cite{Mochizuki} point at the essential role of magnetic fluctuations in both the superconducting pairing mechanism and the electronic transport. The character of fluctuations is closely associated with the magnetic dimensionality and frustration, consequently, capturing the nature of magnetic interactions in this system in the widest possible context appears to be crucial.
 
The Na$_{x}$CoO$_{2}$ system is considered a quasi-2D doped Mott insulator with antiferromagnetic (AF) interaction between the Co atoms on triangular lattice. Sodium between the close-packed CoO$_{2}$ planes serve as a charge reservoir, by donating one electron converts the originally magnetic Co$^{4+}$ (3d$^{5}$) ions with spin \textit{S}=1/2 into nonmagnetic Co$^{3+}$ (3d$^{6}$) ones. The transport and magnetic properties are therefore very sensitive to the doping level and to the degree of frustration, introduced by triangular geometry. The system can be doped over a wide range, and for most \textit{x}, Na$_{x}$CoO$_{2}$ is metallic. However in a narrow range around \textit{x}=0.5, it represents a charge ordered AF insulator \cite{Foo}. Charge ordering as a possible result of frustration and strong correlations has been predicted within several theoretical considerations \cite{Lee, Zhang2} for other doping levels as well. Experimental indications of this process were observed by nuclear magnetic resonance studies \cite{Gavilano}. Phase separation on mesoscopic scale into Co$^{4+}$-rich magnetic domains and Co$^{3+}$-rich regions with itinerant electronic character was strongly indicated by resonance methods \cite{Carretta} and neutron scattering \cite{Helme}. Magnetic phase transition has been observed only for the range 0.7$\le${\textit{x}}$\le${0.95} with clear AF spin density wave resolved for \textit{x}$\ge${0.8} \cite{Bayracki1,Wooldridge}. Theoretical first principle calculations based on local density approximations (LDA) are not conclusive concerning the nature of ground state and type of interactions. More specifically, the LDA calculations \cite{Singh} predict an itinerant ferromagnetic (FM) ground state, whereas LDA+U \cite{Lee} calculations admit also AF ground states. The corresponding calculations for tripled supercells for doping \textit{x}=2/3 lead to solutions favoring AF stacking of FM planes \cite{Johannes}. Analysis of the exchange paths in the system has revealed the feasibility of strong inter-planar superexchange, mediated by by \textit{p}$_{z}$ orbitals of O and considerably facilitated by empty \textit{sp}$^{2}$ hybrid orbitals of Na. Experimentally, although the susceptibility behavior indicated antiferromagnetism, strong in-plane FM fluctuations have been identified by inelastic neutron scattering \cite{Boothroyd}. Consistently with the anisotropic character of the structure in which the inter-plane Co distances are almost 4 times larger than the in-plane ones, the transport was found to be highly anisotropic \cite{Rivadulla} and naturally a quasi-2D magnetic structure has been presumed. Nevertheless, the scheme concerning the magnetic dimensionality is recently being changed: an intense debate brisked up about the possible 3D character of magnetism in this structurally layered system. Recent polarized- and unpolarized-neutron scattering study \cite{Helme, Bayracki2} suggested surprisingly almost isotropic 3D magnetic interactions, specifically, FM exchange within the layers with exchange constant $\sim$ 6 meV and AF correlation between the CoO$_{2}$ layers with $\sim$ 12 meV.

Specific heat on this system has been predominately analysed using the Debye model and the physical interpretations concerned the discussions about the Sommerfeld coefficient \cite{Bayracki1, Wooldridge, Sakurai}. We present an alternative approach, based on mesoscopic phase separation, taking into account the most recent neutron scattering results. The problem of how will the proposed interactions and the existence of phase separation manifest themselves in in the specific heat, is the main subject of the present work. 

Our Na$_{x}$CoO$_{2}$ powder samples with nominal Na content of \textit{x}=0.65, 0.70, 0.75 (hereafter denoted as Na65, Na70 and Na75, respectively) in the doping regime, where magnetic order is about to form, were prepared by the rapid heat-up method \cite{Motohashi}. X-ray powder diffraction confirmed the samples to be single phase of hexagonal $\gamma$-Na$_{x}$CoO$_{2}$ with lattice parameter \textit{a}  around 2.827 \AA~for all samples, while \textit{c}=10.939, 10.907 and 10.892 \AA~for \textit{x}=0.65, 0.70 and 0.75, respectively. Specific heat measurements from 3 K to 360 K have been performed in zero magnetic field and in magnetic field of 9 T using the conventional Quantum Design PPMS-9 equipment.

Fig.1 shows the temperature behavior of specific heat for the Na65 sample, this behavior is similar for all samples. The lattice contribution has been estimated by well established procedure based on harmonic approximation of the phonon spectrum, using both the Debye and Einstein model and involving corrections for anharmonicity \cite{Svoboda, Mihalik}. In the fitting procedure the simple Sommerfeld term \textit{$C_{e}$}=$\gamma.T$ has been used to approximate the electronic part of specific heat, yielding the value of $\gamma$=29 mJ/molK$^{2}$. This value is in good agreement with that reported in the literature \cite{Bayracki1, Wooldridge, Sakurai}. After separation of the electronic and phonon background three features became pronounced in the residual specific heat (Fig.1 right axis):
\begin{figure}[ht]
\includegraphics [width=80 mm]{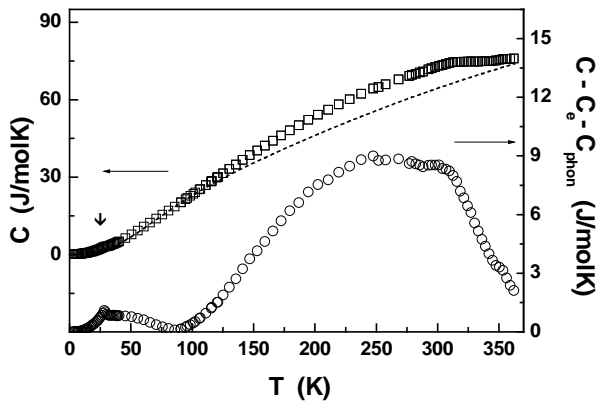}
\caption{\label{fig:epsart} Squares: specific heat experimental data (\textit{C}) for the Na65 sample. Dashed line: fit of the electronic (\textit{$C_{e}$}) and phonon (\textit{$C_{phon}$}) contribution. Circles: residual specific heat after subtracting the electronic and phonon contribution.}
\end{figure}
1) There is a broad maximum around 250 K. According to the literature \cite {Gavilano, Carretta}, this is the temperature range where charge ordering occurs, thus it might be speculated that this bump is a possible indicium of charge ordering in specific heat. The entropy change accompanying this process we found to be proportional to the electron doping. Specific heat measurements on single crystals also have evidenced two sharp peaks in this temperature region \cite {Wooldridge} that could be connected to the anomaly observed in our measurements, considerably broadened because of the powder state of the samples.\\
2) The broad maximum around 40-50 K is attributed to short-range ordering (SRO) in the magnetic domains, as it will be discussed later.\\
3) There is a rather smeared phase transition at \textit{T}$^{*}$$\sim$ 28 K with peak position and shape same in magnetic field of 0 and 9 T (Fig.2). The fact that the transition is not rounded by magnetic field could imply that the main interaction in this compound is AF. This is in agreement with the Curie-Weiss fits of the susceptibility, that revealed a negative Weiss temperature with values around 120 K for all three samples \cite {Zorkovska}. The transition temperature is independent of Na content \textit{x}, on the other hand, the area of the peak is suppressed by doping. This is consistent with the phase separation scenario and suggests that the phase, the transition is related to, does not change, only its volume fraction decreases with doping.\ 

\begin{figure}[ht]
\includegraphics [width=75 mm]{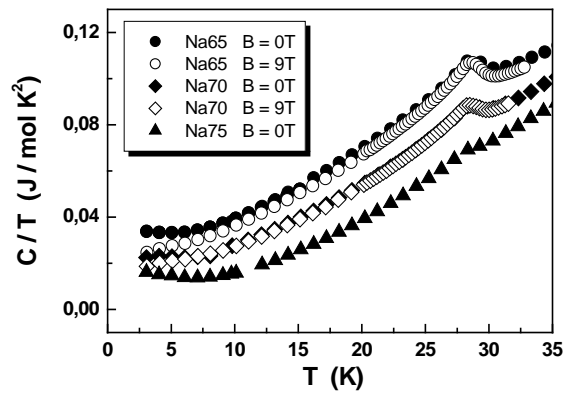}
\caption{\label{fig:epsart}The experimental \textit{C}/\textit{T} vs. \textit{T} data measured in zero magnetic field and in the magnetic field of 9 T. The Na70 and Na75 data are shifted for clarity down by 0.01 and 0.02 J/molK$^{2}$, respectively.}
\end{figure}

Now, let us get closer look at the magnetic part of the specific heat of these unconventional oxides, with the aim to interpret the maximum ascribed to SRO. In this analysis, keeping in mind the suppressing role of Na on the number of magnetic ions, within a simple model only 35\% (30\%, 25\%) of Co atoms in the 65\% (70\%, 75\%) electron-doped system are considered to be magnetic. The volume fraction of magnetic ions of about 21\%, estimated by muon spin rotation in Na$_{0.75}$CoO$_{2}$ \cite{Sugiyama}, shows that the above numbers are reasonable. In the subsequent discussion the specific heat data will be related to one mole of magnetic Co$^{4+}$ ions. 
Assuming phase separation, we examine the applicability of the predictions of localized AF Heisenberg models to the specific heat contribution of the magnetic subsystem (i.e. the domains with localized spins). The choice of Heisenberg model is justified by EPR study and by the obtained value of the g-factor that generally reflects the anisotropy of magnetic ions. More specifically, the estimated value of the g-factor was very close to 2 and the asymmetry of the spectra was found only slight \cite{Carretta}. It should be stressed that the same approach was used in analysis of inelastic neutron scattering data \cite{Helme, Bayracki2}. 

The comparison of the rescaled data to different low-dimensional models can be seen in Fig.3. The first model we tested is the triangular lattice model of spins \textit{S}=1/2 \cite{Elstner}, as the magnetic layers were originally expected to correspond with the structural CoO$_{2}$ layers. The discrepancy between the model and data is considerable. For all samples the qualitatively best agreement is achieved by fitting the quasi-2D square lattice model, involving a weak inter-planar coupling \textit{J}$_{\perp}$ \cite{Sengupta} and this is consistent with the idea of "magnetic layers" perpendicular to the structural ones, as it will be discussed later. However, the obtained parameters should be interpreted with caution, as the model ascribes the peak to long-range ordering (LRO) induced by interplanar exchange coupling. For the obtained value of this parameter the magnetic field of 9 T should induce considerable shift of the peak. The fact, that no shift is observable, suggests that the peak does not reflect LRO mediated by exchange interactions. The estimated in-plane exchange constant, however, is in good agreement with the Weiss temperature obtained from the susceptibility. As the precise exchange paths along the perpendicular axis are not known and on principle the diagonal exchange between the Co ions in the neighboring structural layers cannot be excluded, we tested also the frustrated square lattice model \cite{Bacci}, into which the next nearest interaction \textit{J}$_{2}$ along the diagonal of the plaquette of the lattice is included. The broad maximum is quantitatively better described by this model, the fitting analysis suggests rather strong frustration with \textit{J}$_{2}$ /\textit{J}$_{1}$= 0.7. For completness, also the AF chain model \cite{Blote} is included, this fit is again rather close to the data. \
 
\begin{figure}[ht]
\includegraphics [width=76 mm]{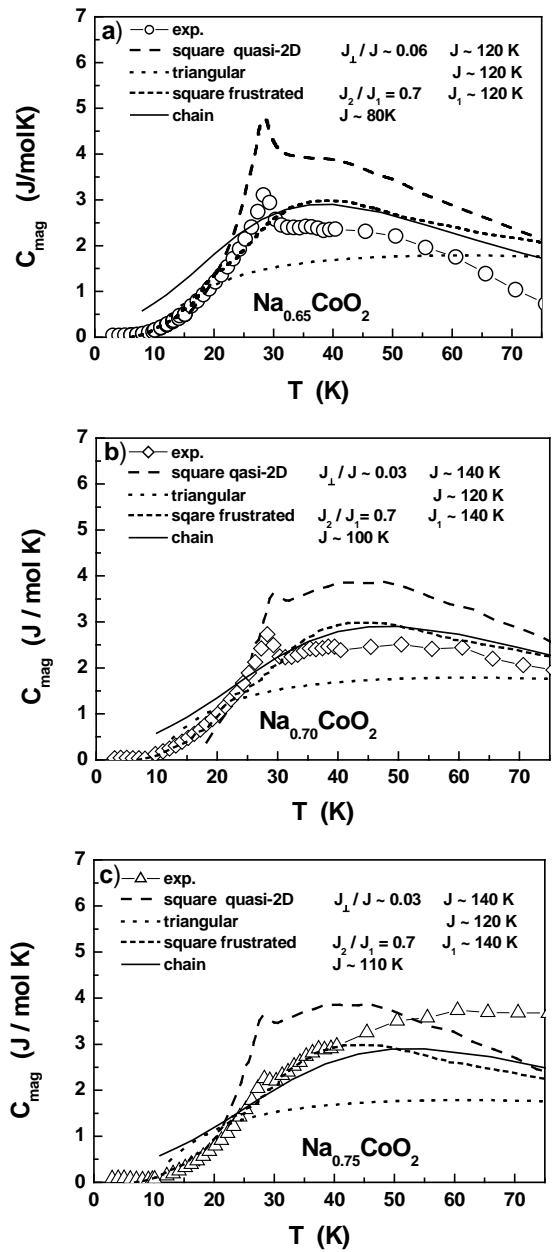}
\caption{\label{fig:epsart}Comparison of the magnetic part of specific heat (symbols) with different low-dimensional AF Heisenberg models.}
\end{figure}

The above analysis indicates that SRO represents a considerable contribution to specific heat and in this way the system does not behave like a 3D magnet. The most interesting result of the above comparisons is the qualitative agreement with the square lattice models and a question arises, whether it is justifiable. Let us put the facts together:\\ 1) The system is separated into mesoscopic domains with localized spins and regions with itinerant electronic states.\\ 2) From the aforecited neutron scattering study of Helme et al. \cite{Helme} follows, that the main AF interaction propagates along the c-axis, perpendicular to the CoO$_{2}$ layers, while FM fluctuations are observed within the layers at 6K. It must be noted, that in this study the localized 3D spin wave approach has been used to analyze the dispersion modes, which does not consider the charge ordering or phase separation in the system. The observed spin excitation spectrum however, hints to the possible phase separation, namely into FM in-plane clusters of Co$^{4+}$ ions in a matrix of non-magnetic Co$^{3+}$. Helme et al. speculate, that in order to obtain consistency with the observed sharp spin modes along the c-axis these clusters would have to be aligned vertically above each other over many layers.\\ 3) The theory of sodium ordering proposed by Zhang et al. \cite{Zhang1} found stable ordering patterns for different dopings in $\gamma$-Na$_{x}$CoO$_{2}$. Regarding the interplay between Na ordering and charge ordering of Co in the planes, considerably weakened in-plane exchange interactions can be expected due to the increased distance between the magnetic Co ions in some specific directions, all the more as the magnetic orbitals are oriented out of plane. In other words, more or less continuous stripe-like regions of magnetic ions are formed within the layers, depending on Na doping level.\\ Summing up these facts, the existence of a low dimensional magnet, most likely in a squared arrangement of magnetic moments with AF interactions perpendicular to the CoO$_{2}$ planes becomes possible. It is noteworthy, that the AF exchange constants obtained from our rough fits (120~-~140~K $\sim$ 10~-~12~meV) are in excellent agreement with that estimated from neutron scattering results \cite{Helme}. Moreover, the increasing tendency of this constant with Na doping supports the reasoning of Johannes et al. \cite{Johannes} about the superexchange enhanced by Na. 
It remains a question, what is the origin of FM correlations. One of the possible explanation could be the itinerant electrons, nevertheless, this question should be investigated further. 

In summary, low temperature specific heat analysis has shown that the dominant contribution to the specific heat in the region of short-range ordering is mediated by quasi-2D antiferromagnetic clusters, perpendicular to the CoO$_{2}$ layers in the Na$_{x}$CoO$_{2}$ (\textit{x}=0.65--0.75) system.  Comparison of the magnetic part of specific heat to the predictions of different AF Heisenberg models suggests, that the  main magnetic interactions are realized on a square lattice rather than on a triangular one. The estimated AF exchange constant is in good agreement with the neutron scattering results of Helme et al. It would be instructive to study the contribution of the other component of the phase separated system, i.e. the itinerant electrons as a possible origin of the observed FM fluctuations that might become more visible below the phase transition.

\begin{acknowledgments}
We thank M. Meisel and P. Carretta for useful comments. This work has been supported by the Grant Agency VEGA, grant No. 1/0430/03, by the Science and Technology Assistance Agency under contracts No. 20-005204 and No. 51-020102. The work of P. Svoboda is part of the research program MSM 0021620834, financed by the Ministry of Education of the Czech Republic. I. Bradari\'{c} and I. Savi\'{c} were supported by the Serbian Ministry of Science, Technology and Development, Grant No. 1899. Material support of U. S. Steel DZ Energetika is gratefully acknowledged.
\end{acknowledgments}

\thebibliography
\end{document}